\documentclass[twocolumn]{revtex4}
\usepackage{graphicx}
\def\ltsim{\vbox {\hbox{\lower 0.9\baselineskip \hbox{$<$}} \break
                 \hbox{\lower 0.2\baselineskip \hbox{$\sim$}} } }

\def\vec#1{{\bf #1}}

\begin{document}
\title{Distinguishing d-wave from highly anisotropic
s-wave superconductors}

\author{L.~S.~Borkowski and P.~J.~Hirschfeld}

\affiliation{
Department of Physics, University of Florida, Gainesville, FL 32611
}

\begin{abstract}
Systematic impurity doping  in the Cu-O plane of
the hole-doped cuprate superconductors may allow one to decide between
unconventional ("d-wave") and anisotropic conventional ("s-wave") states as
possible candidates for the order parameter in these materials.
We show that potential scattering of any strength
always increases the gap minima of such s-wave states,
leading to  activated behavior in temperature with characteristic
impurity concentration dependence in observable quantities such as
the penetration depth.  A magnetic component to the scattering
may destroy the energy gap and give rise to conventional gapless
behavior, or lead to a nonmonotonic dependence of the gap on impurity
concentration.  We discuss how experiments constrain this analysis.
\end{abstract}
\maketitle

%\vskip -.3cm
%\endtitlepage

{\it Introduction.} A number of recent experiments on
hole-doped cuprate superconductors have provided  evidence for a
superconducting state with very large anisotropy, consistent with
actual gap nodes on the Fermi surface.\citep{NMR,photo,lowenergyqps,squid}
The set of experimental results
indicating the existence of low-energy quasiparticle excitations have
been  interpreted in terms of an {\it unconventional}, "d-
wave" pairing state, where we use the term unconventional to mean
that the superconducting order parameter breaks additional
symmetries of the normal state beyond the usual gauge
symmetry.\cite{SigristUeda}  Such an order parameter
${\Delta_k}^d$ has a nontrivial phase variation over the Fermi
surface and changes sign at the node.  Since the quantities measured
in these experiments usually depend on the order parameter only through
the quasiparticle energy $E_k=({\xi_k}^2 +{|\Delta_k}|^2)^{1/2}$, where
$\xi_k$ is the single-particle energy measured relative to the Fermi
level, it is easy to see that an identical result would be obtained by a
measurement on a hypothetical state with order parameter
${\Delta_k}^s=|{\Delta_k}^d|$, which would vanish at the same nodal
points but never change sign.  Since the nodal points in this case are
accidental rather than being enforced by symmetry, it is more
realistic to consider a highly anisotropic s-wave state with very deep
gap minima but no nodes.  Such an order parameter has in fact been
proposed by Chakravarty et al.,\cite{Chakravartyetal} and is quite
difficult to distinguish
from a similar d-wave state if the experiment does not measure
temperatures substantially below the gap minima and is not
sensitive to the gap phase variation.
\vskip -.2cm

As a consequence of these ambiguities, methods of  distinguishing
between unconventional states and anisotropic conventional states
are of great importance.  Josephson tunneling experiments are
sensitive to the order parameter phase and therefore in principle
capable of deciding this
question.\cite{SigristRice,GeshkenbeinLarkin} At present, however,
different Josephson experiments of slightly different concept and
design have reached differing conclusions regarding the order parameter
symmetry\cite{squid,Chaudhari,Dynes}  We have
therefore reexamined the well-studied problem of dirty
superconductors with an eye towards designing further tests which may be
capable of distinguishing d-wave and highly anisotropic s-wave
states.  We find that systematic impurity doping experiments are indeed
sensitive to the
order parameter phase, albeit indirectly, and can provide important
evidence towards the resolution of this question.
%In particular, we
%find that potential scatterers in an s-wave system always give rise to
%activated low-temperature contributions  to the penetration depth and other
%thermodynamic and transport
%quantities, even in the extreme case that the quasiparticle spectrum
%of the analogous {\it pure} d-wave system is identical.  As the
%corresponding prediction for the penetration depth of a dirty d-wave
%superconductor is known to exhibit a $T^2$
%dependence,\cite{Tsquared,felds}  existing experimental data may
%already provide enough evidence to  rule out anisotropic s-wave
%states on this basis.  The primary loophole in this analysis is
%the possiblility that a Zn impurity or other simple defect induces
%a magnetic moment in its vicinity,\cite{Poilblanc} leading to gapless
%behavior with a nonzero residual density of states mimicing
%the "dirty d-wave" picture.   An estimate of the spin-flip time
%in Zn-doped YBCO is used to argue that potential scattering is still dominant.
%\vskip .4cm

{\it Model---Potential Scattering.}   For illustration's sake we consider
a $d_{x^2-y^2}$ state over a cylindrical Fermi surface, $\Delta_k=\Delta_0
\cos 2 \phi$, and a hypothetical s-wave state $|\Delta_0\cos 2\phi|$.
Norman\cite{Norman} has shown that weak potential scatterers eliminate the
nodes
in the s-wave case at the points $\phi=\pi/4, 3\pi/4,....$, increasing
the gap in these directions monotonically with impurity
concentration.   It was  in fact argued in \cite{Norman} that
the dependence of angle-resolved photoemission (ARPES) data
on sample aging\cite{photo} may be construed as evidence
for s-wave superconductivity, but there are alternative explanations
peculiar to the ARPES configuration.\cite{photo}
Here we consider these two states further, and
investigate the effects of strong scattering as well as spin scattering,
and try to make predictions for bulk thermodynamic and transport experiments.
%\vskip .2cm

The properties of a $d_{x^2-y^2}$ state in the presence of elastic impurity
scattering have been extensively investigated in recent
months,\cite{felds,dx2y2thy} but  are in
fact generic to states with lines of nodes on the Fermi surface in 3D
(point nodes in 2D) investigated in the context of heavy fermion
superconductivity.\cite{Gorkov,heavyfs}
An infinitesimal number of impurities suffice
to make the density of states at the Fermi level
nonzero,\cite{Gorkov} giving rise at low temperatures $T\ll
T_c$ to contributions which vary with temperature as their normal
state analogues, but with a smaller prefactor which scales with
impurity concentration.   The penetration depth, which does not have
a normal state analogue but varies as $\sim T/\Delta_0$ in the pure
d-wave state, is known to cross over to a $T^2$ behavior in this so-
called "gapless" regime.\cite{Tsquared,felds}
At the same time, the actual energy gap in the angle-resolved density of
states remains zero along the nodal directions.   All these
characteristics may be understood as consequences of the exact
vanishing of the anomalous impurity self-energy which occurs in
most---but not all---unconventional states.
%\vskip .2cm

The essential differences between s- and d-wave states may be
understood by examining the  single particle matrix
propagator $\underline g$ averaged over  impurity positions, given
by\cite{AG}
\begin{equation}
\underline g(\vec k,\omega_n) = (\tilde\omega\underline\tau^0 +
\tilde\xi_k
\underline\tau^3 + \tilde\Delta_k\underline\tau^1)
%\over
(\tilde\omega^2 -
\tilde\xi_k^2
-\mid{\tilde\Delta}_k\mid^2)
\end{equation}
where the $\underline\tau^i$ are the Pauli matrices and
$\tilde\Delta_k$
is assumed to be a unitary order parameter of s- or d-wave type
in particle-hole and spin
space.
The propagator (1) has the form of the propagator for the pure
system with renormalized frequency $\tilde\omega =
\omega-\Sigma_0(\omega)$,
single-particle energy $\tilde\xi_k = \xi_k + \Sigma_3(\omega)$,
and
order parameter $\tilde\Delta_k = \Delta_k + \Sigma_1(\omega)$,
where the self-energy
due to s-wave impurity scattering has been written
$\underline\Sigma =
\Sigma_i\underline\tau^i$.    For the  particle-hole symmetric systems we
consider here, renormalization of the single-particle energies  can
be important for arbitrary scattering strengths, but are small in
either
the weak {\it or} strong scattering limit.
We will neglect them in what
follows.
%\vskip .2cm

As alluded to above, in odd-parity states and states with certain
reflection symmetries like $d_{x^2-y^2}$,  the off-diagonal
self-energy $\Sigma_1$ vanishes identically and the gap is
unrenormalized, (${\tilde\Delta}_k=\Delta_k$).
Potential scatterers are then
pairbreaking, in "violation" of Anderson's theorem,\cite{Anderson}
but the angular (e.g.,
nodal) {\it structure} of the gap is not changed.   By contrast, in the
anisotropic s-wave case the order parameter $\Delta_k$ is {\it always}
renormalized by a  positive shift which is independent of
$\bf k$ in the s-wave scattering
approximation.  This leads to a smearing of the energy gap anisotropy
leading eventually to an asymptotically isotropic gap in the dirty limit,
as implied by Anderson\cite{Anderson} and calculated explicitly by
various authors.\cite{dirtysc}
%\vskip .2cm

In the absence of $\xi_k$ renormalizations,
the self-energies are given in a
$t$-matrix approximation by
$\Sigma_0={\Gamma G_0}/D;
    ~\Sigma_1={\Gamma G_1}/D$, where  $\Gamma\equiv n_i/(\pi N_0)$ is a
scattering
rate parameter depending only on the concentration of defects $n_i$
and the density of states at the Fermi
energy,
$N_0$, while the strength of a single  scattering  is
characterized
by the cotangent of the scattering phase shift,
$c$.   Here $D\equiv c^2+{G_1}^2-{G_0}^2$ is the denominator
determining the bound state spectrum, and the  $G_\alpha\equiv(1/2\pi
N_0)\Sigma_k
Tr[\underline\tau^\alpha \underline g]$ are components of the integrated,
disorder-averaged propagator. The Born limit corresponds to
$ c \gg 1$, so that $\Gamma/c^2 \simeq \Gamma_N\equiv\Gamma/(1+c^2)$,
where $\Gamma_N$ is the scattering rate in the normal state due to
impurities.   The unitarity or strong scattering limit corresponds to $c=0$.
%\vskip .2cm

{\it Order parameter, critical temperature, and energy gap}.  We first solve
the Dyson equation for the renormalized propagator (1)
together with the gap equation,
%.  The order parameter $\Delta_k$
%is related as usual to the off-diagonal propagator as
$\Delta(k) =  T\sum_n\sum_{k^\prime} V_{kk^\prime}
{\rm Tr} (\tau_1 / 2) \underline g (k^\prime,\omega_n)$,  where $V_{k
k^\prime}\equiv V_{d,s} \Phi_{d,s}(\hat k) \Phi_{d,s}(\hat k^\prime)$ is the
phenomenological pair interaction assumed. 
The order parameter is $\Delta_k={\Delta_0}^{d,s} \Phi_{d,s}(\hat k)$,  with $\Phi_{d,s}=\cos 2 \phi, |\cos 2\phi|$ for $d$ and $s$ wave, respectively.
%In the d-wave case,
%$\Delta_k={\Delta_0}^d \Phi_d(\hat k)$,  with $\Phi_d=\cos 2 \phi$,
%yielding a simple equation for the gap maximum ${\Delta_0}^d$,
%\begin{equation}
%{\lambda_d}^{-1}=\Big\langle{\int}_0^{\omega_D} d\omega
%\tanh {{\beta \omega}\over 2} ~{\rm Re}
%{{\Phi_d^2}\over{\sqrt{\tilde\omega^2-{{\Delta_k}^2}}}}\Big\rangle,
%\end{equation}
%where
%$\lambda_d = V_d N_0$,  and $\langle ...\rangle $ represents
%an angular average over the cylindrical Fermi surface.
%In the s-wave case, on the other hand, it is convenient to put
%$\Delta_k = \Delta_{avg} + \Delta_{1k}$, where $\Delta_{avg}=
%\langle \Delta_k \rangle$ is the gap average over the Fermi surface.
%When impurities are added to the system, it is easy to check that
%${\tilde\Delta}_k=\tilde\Delta_{avg}+\Delta_{1k},$ determined by
%\begin{equation}
%{\lambda_s}^{-1}=\Big\langle{\int}_0^{\omega_D} d\omega \tanh
%{{\beta \omega}\over 2} ~{\rm Re}
%{{\Phi_s\tilde\Delta_k/\Delta_{0}}\over{\sqrt{\tilde\omega^2-
%{\tilde\Delta_k}^2}}}\Big\rangle.
%\end{equation}
%Note this is effectively an equation for $\tilde\Delta_{avg}(\omega)$
%since the angular variation $\Delta_{1k}$ is given.
The initial slope of $T_c$ suppression, $dT_c/d\Gamma_N=-\chi\pi/4 $,
where $\chi\equiv[\langle{\Phi_s}^2\rangle-
\langle{\Phi_s}\rangle^2]/\langle{\Phi_s}\rangle^2$ is $1-8/\pi^2$
for the s-wave and 1 for the d-wave state considered.
In the d-wave case the critical temperature continues to drop rapidly to
zero at a critical concentration of $n_i^c=\pi^2N_0T_{c0}/2e^\gamma$,
whereas the decrease becomes more gradual as
the gap is smeared out in the s-wave case,  finally
varying\cite{dirtysc} as $T_c\sim T_{c0}
[1-\chi\ln(1.154\Gamma_N/\pi T_{c0})]$.
%\vskip .4cm
\begin{figure}[th]
\includegraphics[width=0.27\textwidth]{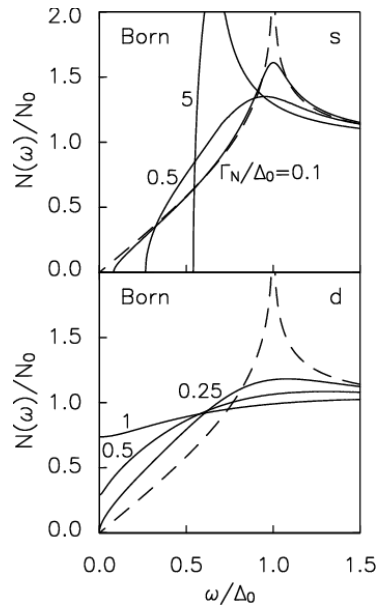}
\caption{Normalized density of states $N(\omega)/N_0$ for s-- and
d--wave order parameters vs. reduced frequency
$\omega/\Delta_0$,shown for various potential scattering rates
$\Gamma_N/ \Delta_0$  in Born approximation.}
\label{fig:1}
\end{figure}

\begin{figure}[th]
\includegraphics[width=0.27\textwidth]{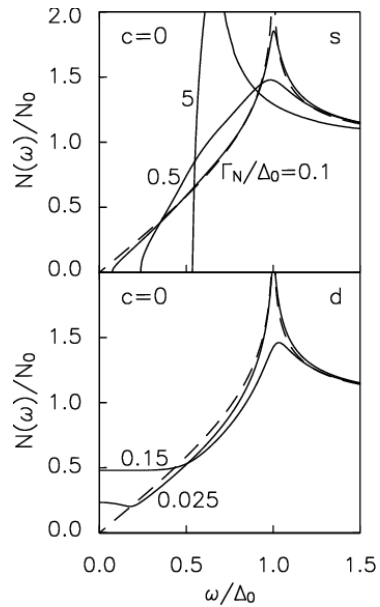}
\caption{Normalized density of states $N(\omega)/N_0$ for s-- and
d--wave order parameters vs. reduced frequency
$\omega/ \Delta_0$, shown for various potential scattering rates
$\Gamma/\Delta_0$  in unitarity limit, $c=0$.}
\label{fig:2}
\end{figure}

It is important to recognize that the
renormalized order parameter ${\tilde\Delta}_k$ in the s-wave case
is only indirectly related to the actual energy gap $\Omega_G$ in the
system, given by
the maximum frequency $\omega$ such that the angle-resolved
density of states $N({\bf k},\omega)\equiv{\rm Im ~ Tr ~[{\underline g}({\bf k},
\omega})]/\pi=0$ for all $\bf k$.  A simple estimate shows that for
small scattering rates, $\Omega_G \sim \Gamma$($\Gamma_N$ in Born limit).
In the dirty limit $\Gamma \rightarrow \infty$, the s-wave superconductor
becomes isotropic with a BCS density of states
$N(\omega) = {\rm Re} ~ [\omega/(\omega^2 - {\Delta_{avg})}^2)^{1/2}$,
as shown in Fig. 1.  In contrast to a d-wave
superconductor,  the self-energies
obtained in the Born approximation and in the resonant scattering limit
are almost equivalent in the highly anisotropic s-wave system.  This
insensitivity to larger phase
shifts arises because of  off-diagonal self-energy corrections which
prevent the occurence of poles in  the t-matrix,  $c^2-{G_0}^2+{G_1}^2
\simeq {\cal O}(1)$ for all $c\ltsim$.  Densities of states for both types of
states in the limit
of resonant scattering are shown in Fig. 2.

%\vskip .2cm
{\it London penetration depth.}  The opening of the energy gap with
increasing impurity concentration is an indelible signature of s-wave
superconductivity.  It will obviously give rise to activated behavior
for $T\ll \Omega_G$ in a wide range of thermodynamic properties,
of  which we have chosen to discuss only one for purposes of illustration,
the temperature-dependent magnetic penetration depth.  For the
model states and Fermi surface under consideration, this may be
expressed as
%\begin{equation}
$%\Big(
[{\lambda_0/\lambda (T)}]^2
%\Big)^2
= \int d\omega ~
{\rm tanh}
%\, 
{{\beta\omega}/2}\,\int {d\phi/ 2\pi}
%\,\,
{\rm Re}~ %\, \Big\{
{{{\tilde\Delta}_k^2}/{({\tilde\omega}^2-
{\tilde\Delta}_k^2)^{3/2}}} $
%\Big\},
%%\end{equation}
where $\lambda_0 $ is the pure London result at $T=0$.
The penetration depth in a d-wave superconductor (Fig. 3, bottom half)
is known to vary
as $\lambda (T)\simeq {\tilde \lambda}_0 + c_2T^2$ at the lowest
temperatures,\cite{Tsquared,felds}  over a
temperature range which widens with
increasing impurity concentration.   The coefficient $c_2$ decreases,
as ${\Gamma}^{-1}$ in the Born limit and $\Gamma^{-1/2}$ in the
resonant scattering case.
The corresponding activated
behavior in the anisotropic s-wave case is easy to distinguish from
the d-wave case when plotted against $(T/T_c)^2$ as also shown in Fig. 3.
The important experimentally relevant signature is of course not simply
the exponential behavior, but the increase in the activation gap with
impurity concentration.
\begin{figure}[th]
\includegraphics[width=0.27\textwidth]{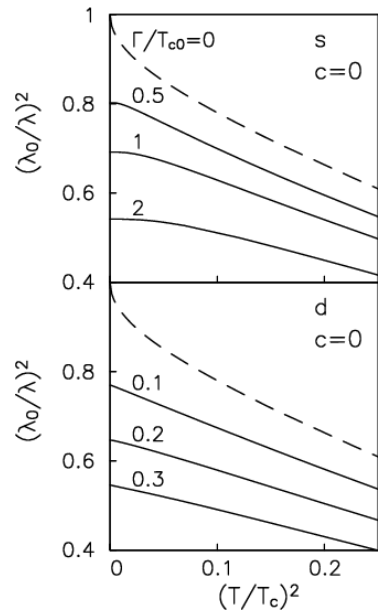}
\caption{Temperature dependence of  normalized magnetic
penetration depth $(\lambda_0/\lambda(T))^2$ for s-- and
d-- wave order parameters vs. reduced temperature $(T/T_c)^2$,
shown for various potential scattering rates
$\Gamma/T_{c0}$  in unitarity limit, $c=0$.}
\label{fig:3}
\end{figure}
\begin{figure}[th]
\includegraphics[width=0.35\textwidth]{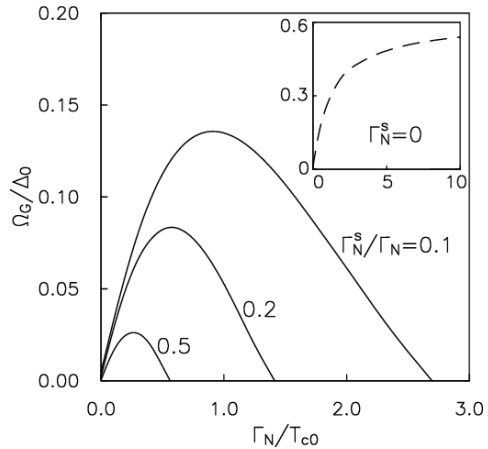}
\caption{Induced energy gap normalized to clean gap maximum,
$\Omega_G/\Delta_0$,  vs. potential scattering rate
$\Gamma_N/T_{c0}$ for different ratios,
$\Gamma_N^s/\Gamma_N$ of magnetic to potential scattering rates.}
\label{fig:4}
\end{figure}
%\vskip .2cm

{\it Spin scattering.}  A simple defect like a vacancy or Zn ion in
the ${\rm CuO_2}$ plane may not behave simply as a potential scatterer, as
assumed above. In the presence of large  local Coulomb interactions,
a magnetic moment may form around the defect site, giving rise to
spin-flip scattering of conduction electrons.\cite{Poilblanc}  This poses
the most
serious obstacle for the direct application of the principle
distinguishing d-wave from anisotropic s-wave systems outlined
above, since magnetic scattering will lead to gapless
superconductivity as in the usual Abrikosov-Gor'kov theory.
Furthermore, even if a gap remains, strong spin-flip scattering may
lead to bound states within
it\cite{oldboundstates,Shiba}
which may  give rise under the proper circumstances to a residual
density of states $N(\omega\rightarrow 0)$ as in the d-wave case.
Here we investigate the competition between the opening of the
energy gap in the s-wave state due to potential scattering and
gapless behavior due to magnetic scattering.
 To this end we
add a term $J {\bf S \cdot \sigma}$ to the Hamiltonian, where $\bf
S$ is a classical spin representing the impurity and $\bf \sigma$ is
the conduction electron spin density, and study the system in an
average  t-matrix approximation analogous to the one applied to the
pure potential scattering case.   The self-energies found in the
presence of both types of scattering reduce in the isotropic s-wave
case to those given by Shiba,\cite{Shiba} but are complicated and
will be given elsewhere.  We find that until the dimensionless
exchange $JN_0$ becomes of ${\cal O} (1)$, the results for the s-wave
system are very similar to those obtained in the simpler Born
approximation, as discussed above.  In this case,
$\Sigma_0=(\Gamma_N + \Gamma_N^s)G_0 $ and
$\Sigma_1=(-\Gamma_N  + \Gamma_N^s) G_1$,\cite{AG}  where
$\Gamma_N^s\equiv n_i J^2 S(S+1)\pi N_0$.  The induced
gap, $\Omega_G$, in the s-wave system may then be shown to vary
as $\Omega_G\simeq \Gamma_N-\Gamma_N^s \ge 0$, but the effects
of self-consistency rapidly become important as the concentration is
increased.  In Fig. 4, we plot $\Omega_G$ as a function of the
impurity concentration through the parameter $\Gamma_N$ for
various assumptions about the scattering character of the impurity
ion, where the quantity $\Gamma_N^s/\Gamma_N $ specifies the
relative amount of magnetic scattering.  The destruction of
the induced gap takes place because the system becomes insensitive
to large amounts of potential scattering, but magnetic impurities
continue to break pairs even at large concentrations.
The gap is  nevertheless found to persist
into the very dirty limit even for systems where the magnetic
scattering is nearly as strong as the potential scattering.
%\vskip .2cm

For weak spin scattering, the bound state in the t-matrix
approximation is found to lie at $\omega\gg \Omega_G$, just below
the {\it average} order parameter $\Delta_{avg}$ deep in the
continuum, and thus plays no role.  Stronger spin scattering does not
change this qualitative behavior at low concentrations until
$J N_0 \simeq 1$  when the bound state lies at the Fermi level
in the classical spin approximation.\cite{Shiba}  In this case the
Kondo effect, neglected here, also becomes important.  It is known
from other analyses\cite{Kondo} that the bound state lies near the
Fermi level, and will therefore give rise to a residual density of
states $N(\omega \rightarrow 0)$, only when $T_K\simeq T_c$.  For
any other ratio of $T_K/T_c$, the bound state will lie at an energy
corresponding to an appreciable fraction of the {\it average} gap in the
system, and hence be irrelevant for our purposes.
%\vskip .2cm

Clearly a quantitative estimate of the relative size of $\Gamma$ and
$\Gamma_N^s$ is required to decide whether spin scattering plays a
role in real high-$T_c$ materials with simple defects.  Walstedt and
co-workers estimated $J N_0  \simeq 0.015$ for a
Zn ion in YBCO, implying that Zn is a nearly pure potential scatterer
in this  system.\cite{Walstedt}  On the other hand,  Mahajan et
al.\cite{Alloul} estimate  $J N_0  \simeq 0.45$.   For a 1\%
Zn concentration,  a magnetic moment of 0.36 $\mu_B$ for Zn in fully
oxygenated YBCO\cite{Alloul} and a density
of states of $1.5/eV$\cite{Alloul}, we find $\Gamma_N^s\simeq
1\times
10^{-4}$ eV.   From the residual resistivities of Zn-doped YBCO
crystals,\cite{Ong} we estimate that a 1\% Zn sample corresponds to
a {\it total impurity} scattering rate of $\Gamma_N^{imp}\simeq
1\times 2\times 10^{-3} eV$, assuming that the inelastic and elastic
contributions to the scattering rate add incoherently.  This suggests
that  potential scattering must dominate the total elastic rate,
$\Gamma_N^s\ll \Gamma$.  On the other hand, the large value of $J
N_0  \simeq 0.45$ deduced for a Zn ion\cite{Alloul} means that the
Kondo effect may be important, and that we cannot completely rule
out the possibility that a bound state sits very close to the Fermi
level.
%\vskip .2cm

{\it Conclusions.}  There is by now a considerable body of
experimental data supporting the picture of  gapless
superconductivity in the cuprate high-$T_c$ materials, with a
residual density of states and low-temperature behavior varying
qualitatively according to the d-wave plus resonant scattering
model.\cite{NMR,Bonnhardybigpaper}  This data stands in
apparent
contradiction to the well-known effect of small amounts of potential
scatterers on anisotropic s-wave superconductors, namely the
smearing of energy gap anisotropy.  This continues to hold even for
extremely anisotropic systems with nodes, as illustrated by the
simple theory presented here for a representative order parameter.
We believe that this data strongly suggests that the pairing is
unconventional in these materials, but the above analysis does not
as it stands allow one to distinguish among possible candidate
unconventional
states  (e.g,, $d_{x^2-y^2}$ and $d_{xz}$) without further quantitative
comparison.  It should be noted that time-reversal breaking
unconventional states with a
gap will become gapless in the presence of pure
potential scattering.
%\vskip .2cm

As we have briefly discussed, the major difficulty inherent in such
an analysis is the possibility that even an apparently "inert" impurity
such as Zn or a vacancy in the Cu-O planes may induce local spin
correlations in the strongly interacting electron system, leading to
spin-flip scattering.  Ruling out  gapless superconductivity induced
by magnetic scattering then becomes a quantitative problem.
Gapless behavior in films suggests that a resonant scattering
mechanism of some type must be present in order to induce a
significant residual density of states with comparatively little $T_c$
suppression.  We have shown, however,  that resonant potential
scattering does not take place in s-wave systems, and argued that
low-energy resonant spin scattering is much less likely than in the
isotropic case.
We have furthermore made a crude estimate of the importance of
spin-flip scattering in Zn-doped YBCO crystals which indicates
these materials are dominated by potential scattering and should therefore
exhibit an induced gap if the superconducting state is s-wave.
%\vskip .2cm
%\smallskip

We are grateful to N. Goldenfeld for useful discussions and encouragement.
Some numerical computations were performed on the Cray-YMP at the Florida
State University Computing Center.

\end{document}